\documentclass[pra,twocolumn,superscriptaddress,nobalancelastpage]{revtex4-1}

\usepackage{amsmath,stmaryrd,xcolor,graphicx,amssymb}
\usepackage[binary-units=true,detect-weight=true,detect-family=true]{siunitx}
\definecolor{darkred}{rgb}{0.5,0,0}
\definecolor{darkgreen}{rgb}{0,0.5,0}
\definecolor{darkblue}{rgb}{0,0,0.5}
\usepackage[colorlinks,breaklinks,linkcolor=darkred,citecolor=darkgreen,urlcolor=darkblue]{hyperref}
\newcommand{\me}{\mathrm{e}}
\newcommand{\red}[1]{\textcolor{black}{#1}}

\begin{document}

\title{Experimental relativistic zero-knowledge proofs}
\author{Pouriya Alikhani}
\affiliation{School of Computer Science, McGill University, Montr\'eal, Qu\'ebec, Canada}
\author{Nicolas Brunner}
\affiliation{Department of Applied Physics, University of Geneva, 1211 Gen\`eve, Switzerland}
\author{Claude Cr\'epeau}
\affiliation{School of Computer Science, McGill University, Montr\'eal, Qu\'ebec, Canada}
\author{S\'ebastien~Designolle}
\affiliation{Department of Applied Physics, University of Geneva, 1211 Gen\`eve, Switzerland}
\author{Rapha\"el Houlmann}
\affiliation{Department of Applied Physics, University of Geneva, 1211 Gen\`eve, Switzerland}
\author{Weixu Shi}
\affiliation{Department of Applied Physics, University of Geneva, 1211 Gen\`eve, Switzerland}
\affiliation{Department of Electronic Science, National University of Defense Technology, 410073 Changsha, China}
\author{Nan Yang}
\affiliation{\mbox{Department of Computer Science \& Software Engineering, Concordia University, Montr\'eal, Qu\'ebec, Canada}}
\author{Hugo Zbinden}
\affiliation{Department of Applied Physics, University of Geneva, 1211 Gen\`eve, Switzerland}
\date{12 July 2021}

\begin{abstract}
  Protecting secrets is a key challenge in our contemporary information-based era.
  In common situations, however, revealing secrets appears unavoidable, for instance, when identifying oneself in a bank to retrieve money.
  In turn, this may have highly undesirable consequences in the unlikely, yet not unrealistic, case where the bank's security gets compromised.
  This naturally raises the question of whether disclosing secrets is fundamentally necessary for identifying oneself, or more generally for proving a statement to be correct.
  Developments in computer science provide an elegant solution via the concept of zero-knowledge proofs: a prover can convince a verifier of the validity~of~a~certain statement without facilitating the elaboration of a proof at all.
  In this work, we report the experimental realisation of such a zero-knowledge protocol involving two separated verifier-prover pairs.
  Security is enforced via the physical principle of special relativity, and no computational assumption (such as the existence of one-way functions) is required.
  Our implementation exclusively relies on off-the-shelf equipment and works at both short (\SI{60}{\meter}) and long distances ($\geqslant$\SI{400}{\meter}) in about one second.
  This demonstrates the practical potential of multi-prover zero-knowledge protocols, promising for identification tasks and blockchain applications such as cryptocurrencies or~smart~contracts.
\end{abstract}

\maketitle

\textit{Introduction.---}
In a foreign city where you know absolutely no one, you go to an automatic teller machine to obtain a handful of local cash.
You have never heard of the bank owning that teller machine, yet when requested for your Personal Identification Number to obtain money you blindly provide it.
No joke, you give away that super unique information to a complete stranger.
But why?
Because of the cash you get in return?
There is actually zero solid reason to trust that teller machine.
You should never have to give away this private information to anyone at all!
But how could we prove who we are without giving away such a secret piece of data?

\begin{figure*}[ht!]
  \centering
  \includegraphics[width=\textwidth]{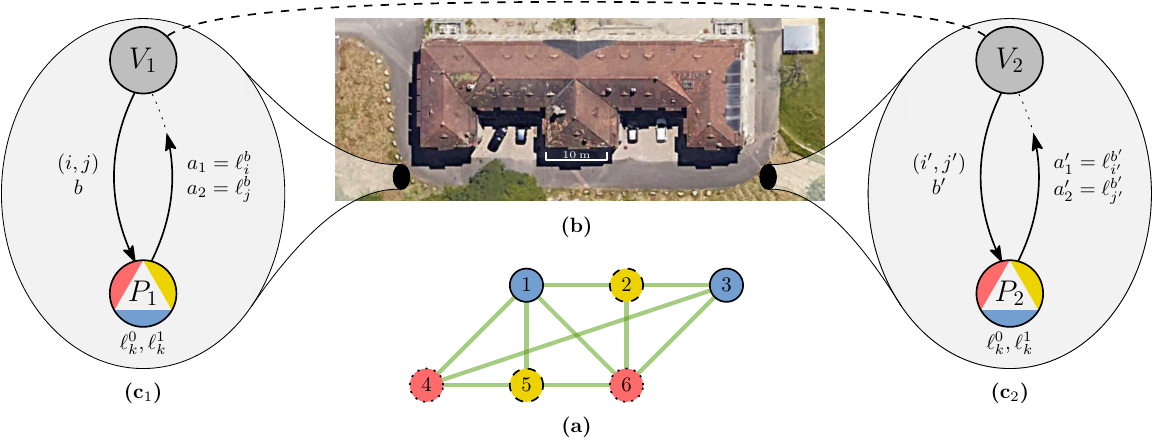}
  \caption{
    \textbf{Relativistic zero-knowledge protocol for three-colourability on a short distance.}
    Two separated provers try to convince a verifier that they know a given graph is three-colourable without facilitating the elaboration of a three-colouring.
    \textbf{(a)}~A three-colourable graph with six vertices and ten edges.
    The three-colouring depicted here is such that $c_1=c_3=0$~(full blue), $c_2=c_5=1$~(dashed yellow), $c_4=c_6=2$~(dotted red); vertices linked by an edge are indeed of different colours.
    \textbf{(b)}~Satellite view~\cite{Google} of the building of the experiment.
    The distance between the two parties involved is \SI{60}{\meter}, that is, \SI{200}{\nano\second} at the speed of light.
    This separation makes the communication between the two provers impossible at this time scale due to the nonsignalling principle of special relativity.
    The verifiers simultaneously trigger their questions to their provers by means of an optical fibre (dashed line).
    \textbf{(c)}~Illustration of a round of the protocol on both verifier-prover pairs.
    Each verifier sends (downward arrow) an edge and a bit $b$ to their prover, who should answer (upward arrow, incomplete to emphasise the chronology) their $b$th labellings at the end points of the edge: for all vertex $k$ the provers have indeed pre-agreed on two labellings $\ell_k^0,\ell_k^1\in\{0,1,2\}$ that should sum up to a three-colouring, namely, ${\ell_k^0+\ell_k^1\equiv c_k\pmod3}$.
    When asking the same edge on both sides and opposite bits, the verifiers can check, thanks to the definition of the labellings, that the provers know that the graph is three-colourable.
    To make sure that the provers are not cheating the verifiers can also send the same bit with edges sharing (at least) one vertex; the consistency of the provers' answers can then be tested.
    By repeating this procedure many times the verifiers can make the probability for dishonest provers to pass the protocol arbitrarily small (soundness).
    However, even with all the provers' answers in hand, the verifiers are not more efficient at elaborating a three-colouring than initially (zero-knowledge).
  }
  \vspace{-0.4cm}
  \label{fig:zkpfig}
\end{figure*}

The idea behind zero-knowledge proofs was born in the middle of the 1980's~\cite{GMR85,GMR89} and formalises the possibility to demonstrate knowledge of a secret information without divulging it.
A natural application is the task of identification, where a user can demonstrate their identity via the knowledge of a secret proof of a mathematical statement they created and published.
A well-known example is the RSA cryptosystem~\cite{RSA78} in which the mathematical secret is the factorisation into two huge prime numbers of an even larger number.
In this work we consider the problem of three-colouring of graphs: an instance is a graph (nodes and edges attaching some of them to one another) and a proof of three-colourability assigns to each vertex one out of three possible colours in a way that any two vertices connected by an edge have different colours, see Fig.~\ref{fig:zkpfig}(a).
Some graphs are three-colourable, some are not, and the general problem of deciding whether a graph is three-colourable has no known efficient solution.
However, given a colouring it is extremely easy to efficiently check if it is \emph{proper}, i.e., whether the end points of every edge are assigned different colours.
For this reason, three-colourability is a problem in $\mathbf{NP}$, the class of all problems that are efficiently verifiable given a solution~\cite{GJ79}.
Moreover, it is also $\mathbf{NP}$-complete because an instance of any problem in $\mathbf{NP}$ can be efficiently simulated by an instance of three-colourability, so that if this latter were in $\mathbf{P}$, the class of all problems efficiently solvable, then we would have $\mathbf{P} = \mathbf{NP}$, an equality which has been the most famous challenge of theoretical computer science for the last half century and which remains unsolved.

A zero-knowledge proof for three-colourability has been introduced in Ref.~\cite{GMW91} by assuming the existence of one-way functions, that is, functions that can be efficiently computed but for which finding a preimage of a particular output cannot.
The zero-knowledge proof guarantees that upon participation to such an interaction, a prover would convince a verifier of the validity of the statement when it is indeed valid (completeness), would not convince the verifier when it is invalid (soundness), while not allowing the latter to improve their ability to find a proper three-colouring (zero-knowledge), but this is under the assumption that one-way functions exist.
It is widely believed that a zero-knowledge proof for any {$\mathbf{NP}$-complete} problem such as three-colourability is not possible without this extra computational assumption.
If not, this would lead to vast implications in the world of complexity~\cite{For87}.
However, this feature is generally undesirable as it significantly weakens the long-term security of such zero-knowledge protocols, which are used, e.g., in certain crypto-currencies~\cite{BCG+14}.
This may have important consequences, as security would be fully compromised if the specific one-way function used in the protocol is (later) found to be efficiently invertible.
This aspect is particularly relevant given recent advances on quantum computing~\cite{BL17,AAB+19}.

Remarkably, it is possible to devise zero-knowledge protocols without the need of any computational assumption.
The key idea, as developed by Ben-Or, Goldwasser, Kilian and Wigderson~\cite{BGKW88}, is to generalise the interactive proof model such that \emph{several} provers are now trying to convince a verifier of the three-colourability of a graph in perfect zero-knowledge without the need of any further assumption.
Intuitively, this approach reflects the strategy used by police investigators when interrogating suspects in separate rooms in order to discern the truth more easily: it is harder to collectively lie about the validity of a statement when interrogated {\em separately}.
The key difference between the multi-prover scenario and the original definition of interactive proof rests in the possibility to prevent several provers from talking to each other, a single prover always being able to talk to themself.
This naturally suggests the use of spatial separation to enforce the impossibility to communicate~\cite{Kil90,Ken99}, at least for some short period of time: assuming the principle of special relativity (nothing can signal faster than the speed of light) and sending queries to the different provers simultaneously, there is a short time window during which they are physically unable to signal to each other given they have to respond fast enough to their nearby verifier.
So far, these ideas have been mainly of purely theoretical interest, as known protocols required extremely large information transfer between the provers and verifiers, which prohibited their implementation.

In this work, we report experimental realisations of relativistic zero-knowledge proofs for an $\mathbf{NP}$-complete problem (three-colourability).
Specifically we simplify and develop an efficient implementation of the protocol recently established in Ref.~\cite{CMS+19} for two separated verifier-prover pairs.
In practice, key challenges involve the generation of adequate large three-colourable graphs, as well as an efficient management of the randomness shared between the provers, achieved via suitable error-correcting codes.
We report on two experiments: first, using Global Positioning System (GPS) clocks to synchronise the two verifiers, we performed the protocol at a distance of \SI{400}{\meter}; second, using a triggering fibre between the two verifiers, we conducted the same test at a shorter distance of \SI{60}{\meter}.
In both cases, the full running time was about one second.
The first implementation shows that the protocol at large distances is rather effortless since the wide relativistic separation only demands a moderate speed on the provers' side; the second one demonstrates a clear potential for serviceable applications.
Importantly, the security is enforced by relativistic constraints, and does not rely on any computational hypothesis such as the existence of one-way functions.
Note that the aforementioned $\mathbf{NP}$-completeness guarantees that any application based on a problem in $\mathbf{NP}$ can be (polynomially) cast into an instance of our protocol.
For example, if you trust the Advanced Encryption Standard (AES) as a secure cryptographic primitive, you can transform AES instances into three-colourable graphs.
Our implementation achieves security against classically correlated provers and we discuss the prospects of extending the security to the general case of quantum-mechanically correlated provers below.

\textit{Protocol.---}
We start by presenting the zero-knowledge proof that we used in the experiment.
Let $(V,E)$ be a finite undirected graph, namely, a finite set $V$ of vertices and a collection $E$ of edges, that is, unordered pairs of (distinct) vertices.
We further assume that this graph is three-colourable, see Fig.~\ref{fig:zkpfig}(a).
In the following we denote the three different colours by 0, 1, and 2 and we refer to proper colourings simply as ``colourings'', whereas we call improper ones ``labellings''.

The protocol we implemented is a simplified version of the one presented in Ref.~\cite{CMS+19}.
It is schematised in Fig.~\ref{fig:zkpfig}(c) and fully explained in Methods~\ref{app:protocol}.
In a nutshell, the (honest) provers share in advance two labellings of the graph (summing up to a three-colouring), which they can use to correctly answer the verifiers' questions.
Combining both answers, the verifiers can then be convinced, round after round, that the provers are not cheating, and this without getting any information that they dit not have initially.
With a number of rounds of \red{$5|E|k$}, where $|E|$ is the number of edges in the graph, classically correlated provers can dishonestly pass the protocol with probability at most $\me^{-k}$.

\textit{The graph.---}
From a theoretical perspective, the three-colourability problem is $\mathbf{NP}$-complete.
For the implementation we need a concrete graph together with a three-colouring of it.
Here comes a complication: though the general problem is ``hard'', there exist efficient algorithms in many cases.
In order to overcome this difficulty we use sufficiently large {\em critical} graphs.
A four-critical graph is not three-colourable but is such that the deletion of any edge gives rise to a valid three-colouring for the resulting graph.
Ref.~\cite{MN08} proposes an algorithm to build large critical graphs corresponding to very hard instances of three-colourability, a fact corroborated by extensive experimental evidence.
However, no method for generating a three-colouring on the way is provided therein, so that we adapted the technique to our needs; see Methods~\ref{app:graph}.
The graph used in the following experiments has $|V|=588$ vertices and $|E|=1097$ edges, so that reaching a security parameter of $k=100$, widely considered safe~\cite{KL14}, takes about \red{half a} million rounds.

\textit{Implementation.---}
Our protocol features two separated verifier-prover pairs.
For its implementation, the critical aspect dwells on the speed of the answer on the provers' side; therefore they were operated on field-programmable gate-arrays (FPGA) to reduce the communication latency, speed up the computation, and improve its time reliability.
On the verifiers' side, FPGAs were also used for communication, together with standard computers for global monitoring and checking of the answers; see Methods~\ref{app:hardware} for details of the hardware, which builds upon techniques developed for the implementation of bit commitment~\cite{VMH+16}.

As the protocol involves a significant number of rounds and requires the provers to share in advance some randomness, this resource must be used sparingly.
For instance, it is easy to see that, starting from a shared three-colouring, storing a single permutation of three elements is enough to draw a random three-colouring in each round.
Regarding the remaining shared randomness needed in the protocol, it requires at first sight one random trit per vertex and per round, which is not affordable.
On second thought only four (two per prover) may suffice but for this the provers should know which question their partner was asked, i.e., which trits were ``consumed'', which is not possible.
Drawing a connection with error-correcting codes~\cite{LLH+14,TV09} we could nonetheless overcome this difficulty; see Methods~\ref{app:randomness}.

Note that two timescales are involved in the experiment: the speed of the exchange between the verifiers and the provers and the repetition rate of the rounds.
The former fixes the minimum distance required between the two locations and is limited by the speed of computation on the provers' side; the latter determines the time that the protocol takes to reach a given security parameter.

In the next sections, we explain our implementations of the protocol in two complementary spatial domains.

\textit{Long-distance.---}
The two verifier-prover pairs are placed in different buildings on the campus at \SI{390}{\meter} from one another, corresponding to a time separation of \SI{1.3}{\micro\second}.
The synchronisation relies on GPS clocks as in Ref.~\cite{LKB+15}.
Both verifiers send to their neighbouring prover a stream of challenges at a frequency of \SI{0.3}{\mega\hertz}.
As soon as they receive a challenge the provers compute their answers based on their shared data, see Fig.~\ref{fig:zkpfig}(c).
Taking into account the imprecision of the system used, the total time elapsed between the emission of the verifiers' challenges and the reception of the provers' answers is \SI{840}{\nano\second}, which is below the \SI{1.3}{\micro\second} time separation between the parties, thus fulfilling the soundness requirement.
The whole protocol with \red{half a} million rounds runs in about \red{\SI{2}{\second}}.

Note that the theoretical minimum distance between the verifiers is fixed by the \SI{840}{\nano\second} in which the provers respond and is thus about \SI{250}{\meter}.
Also there is no upper bound for this distance since the two verifier-prover pairs are disconnected in this case.
Applications involving faraway actors may be designed based on this simple protocol~\cite{VMH+16}: as the distance between the verifier-prover pairs increases, the one between the verifier and the prover within a pair becomes less constrained.
Typically verifier-prover pairs widely separated would allow the provers to be anywhere in the verifiers' cities.

\pagebreak

\textit{Short-distance.---}
The two verifier-prover pairs are placed on two tables outside of the university building  at \SI{60}{\meter} from one another, corresponding to a time separation of \SI{200}{\nano\second}, see Fig.~\ref{fig:zkpfig}(b).
A trigger signal is sent from the first verifier to the second who sends, upon receipt, a challenge to \red{their neighbouring} prover.
The first verifier delays the emission of their challenge by the time the trigger will take to be transferred to the second verifier.
So both verifiers send to their neighbouring prover a stream of challenges.
Again, as soon as they receive a challenge the provers compute their answers based on their shared data and send them back to the verifiers.
Altogether a round is achieved in a maximum \SI{192}{\nano\second}, thus constraining the verifier to be at a minimal distance of \SI{57.6}{\meter}; see Methods~\ref{app:hardware} for details.
With a repetition rate of \SI{0.5}{\mega\hertz} the whole protocol with \red{half a} million rounds runs in \red{about \SI{1}{\second}}.
Note that with the hardware used and its time and memory limitations, it would still be possible to gain an order of magnitude in the size of the graphs, namely, $|V|\sim5\times10^3$ and $|E|\sim10^4$, while keeping the same security parameter $k=100$ and a reasonable total time (about ten seconds).
With the improvements mentioned below, this limit could be further pushed to $|V|$ and $|E|$ of the order of $10^5$.

In our implementation the time needed for an exchange between the provers and the verifiers is mostly constrained by the latency of the hardware, primarily the one of the multi-gigabit transceivers used for the optical links.
The computation of the provers' answers (memory look-up and calculation) is done in a single clock cycle (here \SI{8}{\nano\second}).
A parallel communication with dedicated input/outputs could reduce the transfer time from and to the physical pins of the provers' FPGAs, adding no more than another clock cycle of delay, hence bringing the exchange time down to \SI{16}{\nano\second}.
Moreover, implementing the scheme on a state-of-the-art application-specific integrated circuit (ASIC) technology would further reduce the clock cycle, thus the overall delay.
Therefore it seems possible to run a full exchange in only a few nanoseconds so that the two verifier-prover pairs could eventually be placed about a metre away from each other.

\textit{Quantum provers.---}
So far we have only considered the case of classically correlated provers.
However, it would be desirable to extend the security to the case of quantum provers.
This is because they could establish stronger correlations than classically correlated ones, a phenomenon \red{due to quantum entanglement}~\cite{Bel64}.
Concerning our protocol, it is at the moment unknown whether it remains secure against two quantum provers though it appears to be the case.
In principle there also exist protocols that are secure against such quantum provers~\cite{KKM+11,CL17,Ji13} but they are currently unpractical (see Methods~\ref{app:quantum}) because too many rounds are required under existing analysis.
Therefore, in all cases, improving theoretical proofs clearly represents the key challenge.

\textit{Conclusion.---}
We have demonstrated that a relativistic zero-knowledge proof for the $\mathbf{NP}$-complete problem of three-colourability is possible in practice, even for small distances.
For the example mentioned in the introduction, one could thus conceive a teller machine with two separate ports; customers may then simply spread their arms and insert a pair of chips to identify themselves by proving they know their (public) graph is three-colourable.
Given the simplicity of the operations on the provers' side in our protocol, these chips may furthermore be integrated in (two) cell phones.
More generally, these ideas may find applications in a wide range of areas where the concept of zero-knowledge is relevant, such as blockchain systems and smart contracts~\cite{BBHR18}, electronic voting and auctions~\cite{Gro05,MR14}, as well as nuclear warhead verification~\cite{GBG14}.

\textit{Acknowledgments.---}
Financial supports by the Swiss National Science Foundation (Starting grant DIAQ, NCCR-QSIT) and the European project OpenQKD are gratefully acknowledged.
The Montr\'eal team is grateful to Qu\'ebec's FRQNT and Canada's NSERC for making this work financially possible.

\bibliography{ABC+20}
\bibliographystyle{sd2}

\appendix

\onecolumngrid
\vspace{-0,3cm}
\section*{Methods}
\twocolumngrid

\subsection{Protocol}
\label{app:protocol}

In this section we describe the protocol we implemented together with the strategy used by the verifiers, which gives rise to the number of rounds presented in the main text.
All this is adapted from the very similar though more complex protocol presented in Ref.~\cite{CMS+19}.

\begin{figure}[ht!]
  \centering
  \includegraphics[width=\columnwidth]{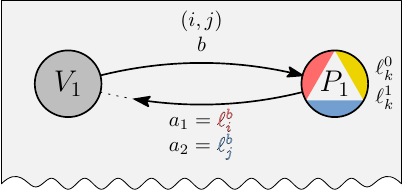}
  \vspace{-0.2cm}
  \caption{
    Illustration of a round of the protocol.
    The colours are consistent with those of Fig.~\ref{fig:zkpfig}(a) and depict a typical round where the verifiers ask the same edge \red{to the provers, here $(1,2)$, but where $b\neq b'$ so that they check} in the end that $a_1+a'_1\not\equiv a_2+a'_2\pmod3$.
    In this example we have $\ell_1^0=2,\ell_1^1=1,\ell_2^0=0,\ell_2^1=1$; note that, despite the adjacency of the vertices 1 and 2, the equality $\ell_1^1=\ell_2^1$ is legal as the labellings $\ell_k^b$ do not need to be colourings.
  }
  \vspace{-0.3cm}
  \includegraphics[width=\columnwidth]{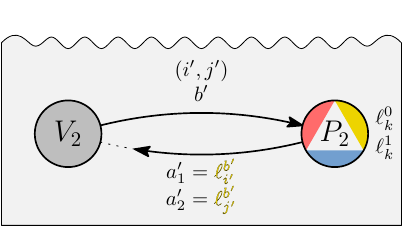}
  \vspace{-0.4cm}
  \label{fig:round}
\end{figure}

The protocol involves two verifiers and two provers.
Initially the two provers pre-agree on random three-colourings $c_k(n)\in\{0,1,2\}$ for $k\in V$ and $n$ identifying the round.
In the following, the dependency in $n$ will be omitted for conciseness.
For all vertex $k$ they also choose two labellings $\ell_k^0$ and $\ell_k^1$ such that the equality ${\ell_k^0+\ell_k^1\equiv c_k \pmod3}$ holds.
Recall that the labellings $\ell^0$ and $\ell^1$ do not need to have different values on adjacent vertices.
A round is then illustrated in Fig.~\ref{fig:round} and consists in (i)~the first, resp.~second, verifier providing their prover with an edge $\{i,j\}\in E$, resp.~$\{i',j'\}$, and a bit $b\in\{0,1\}$, resp.~$b'$, (ii)~the first, resp.~second, prover answering $(\ell_i^b,\ell_j^b)$, resp.~$(\ell_{i'}^{b'},\ell_{j'}^{b'})$, and (iii)~the two verifiers checking the provers' answers as described in the next two paragraphs.
If none of the parties abort the protocol, then we repeat rounds until a certain security level is reached, see below.
The verifiers' tests follow two different paths.

On the one hand, the verifiers can check that the provers do indeed know that the graph is three-colourable.
This test is done when both verifiers send the same random edge ${e=\{i,j\}=\{i',j'\}=e'\in E}$ and when $b\neq b'$.
Then the answers $(a_1,a_2)$ and $(a'_1,a'_2)$ of the two provers are accepted if and only if ${a_1+a'_1\not\equiv a_2+a'_2\pmod3}$.

On the other hand, the verifiers can test the consistency of the provers' answers.
When the edges sent share at least one vertex (say, $i=i'$) and when the bits sent are equal ($b=b'$), then the verifiers accept if and only if the corresponding answers of the two provers are equal ($a_1=a'_1$).
This test prevents the provers from answering in a way that would take no account of the edges asked but would only aim at passing the previous check.

For honest verifiers and honest provers (when the graph is three-colourable), it is easy to see that following the protocol will always lead to acceptance.
This property of the protocol is referred to as \emph{completeness}.

For honest verifiers and dishonest provers (when the graph is not three-colourable), the \emph{soundness} refers to the verifiers being able to reveal the cheat with very high probability when performing many rounds.
Intuitively, if the answers of a prover (say, $P_2$) reach their corresponding verifier ($V_2$) before the question of the other one ($V_1$) could have, by any means, made its way there, then this prover ($P_2$) must have answered without knowing what the other one ($P_1$) has been asked.
By separating the verifier-prover pairs by a sufficient distance and by timing the protocol carefully, we can use the nonsignalling principle of special relativity to create this separation in order to make the protocol sound against classically correlated provers.
We discuss the case of quantum provers in Methods~\ref{app:quantum}.

For dishonest verifiers (trying to get any knowledge of a three-colouring) and honest provers, the \emph{zero-knowledge} property amounts to the verifiers getting no knowledge whatsoever upon interaction with the provers.
The above protocol satisfies this property~\cite{CMS+19}.

From the cases described above, we get the main features of a good strategy for the verifiers to detect cheating provers.
Typically, asking edges with no vertex in common is of no interest and the two tests described above should be somehow balanced.
When we fix the strategy adopted by the provers, the probability for cheating provers to pass one round can be computed and from there the number of rounds required to reach a given security level.
\red{In a similar fashion to Ref.~\cite{CMS+19} we used the subsequent strategy for the verifiers.}
First the edge $\{i,j\}$ and the \red{bit $b$} are chosen (uniformly) at random.
Then with probabilities $\frac15$, $\frac25$, and $\frac25$ (respectively), one of the three following options is chosen: (i) the edges are chosen to be equal and the \red{bits} opposite, that is, $\{i',j'\}=\{i,j\}$ and \red{$b' \neq b$}; (ii) \red{the bits are chosen to be equal and the second edge randomly among those containing $i$, that is, $b'=b$}, $i'=i$, and $\{i',j'\}\in E$; (iii) \red{the bits are chosen to be equal and the second edge randomly among those containing $j$, that is, $b'=b$}, $j'=j$, and $\{i',j'\}\in E$.
With this strategy, when the number of rounds is \red{$5|E|k$}, where $|E|$ is the number of edges in the graph, classically correlated provers can dishonestly pass the soundness tests with probability at most $\me^{-k}$.

\pagebreak

Note that the amount of data exchanged is very small compared to previous protocols: in Ref.~\cite{CL17} this quantity is polynomial in the number $|V|$ of vertices while here it is only logarithmic in $|V|$.
This feature allows for short distances between the verifier-prover pairs since the communication time is short, even for large graphs.

\subsection{Graph generation}
\label{app:graph}

In this section we describe how we construct large three-colourable graphs which are hard to colour \emph{together} with a three-colouring.

In Ref.~\cite{MN08} Mizuno and Nishihara give (i) seven small graphs \red{which} are four-critical, that is, not three-colourable but such that any graph obtained by deleting any edge is three-colourable, and (ii) a procedure to assemble two four-critical graphs into a (bigger) four-critical graph.
Typically, the method consists in replacing one edge of the first graph by the second one.
Importantly, the small and assembled graphs do not contain any near-four-clique, that is, any subgraph with four vertices all connected to one another except for one pair, e.g., $\boxslash$.
Such structures indeed appear as weaknesses exploitable by algorithms looking for a three-colouring and should thus be avoided.
With their procedure they experimentally demonstrated using various softwares that the complexity of the resulting instances was exponential in the number of vertices.

However, they do not include any algorithm to keep track of the three-colourings that arise upon removal of an edge.
We developed such a method to build large graphs that are very hard to three-colour together with a three-colouring.

\begin{figure*}[ht!]
  \centering
  \includegraphics[height=5.7cm]{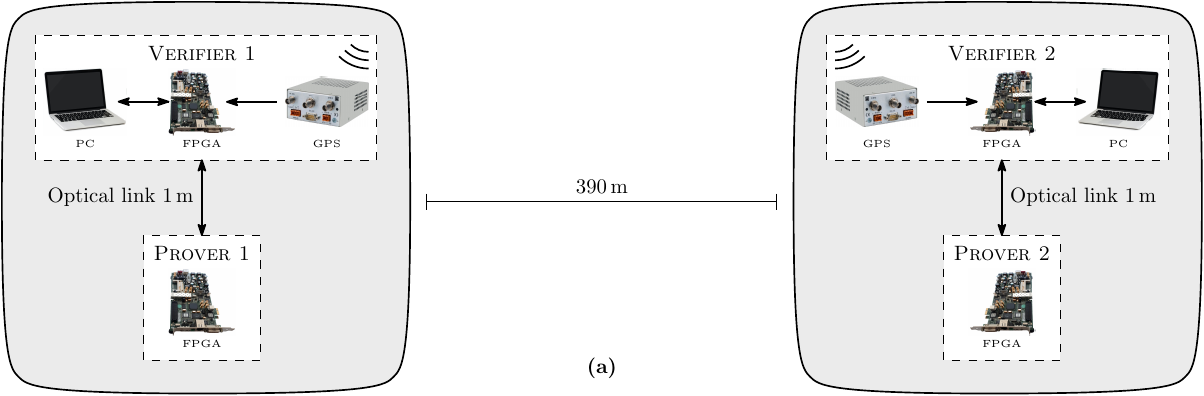}\\
  \vspace{1cm}
  \includegraphics[height=5.7cm]{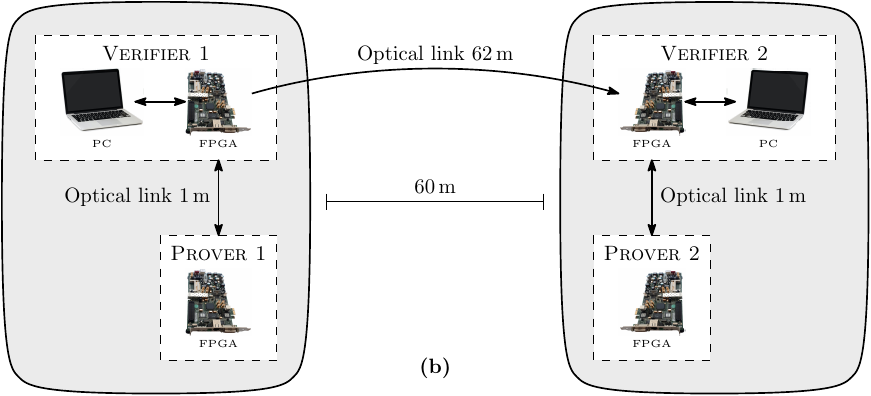}
  \caption{
    Illustration of the hardware used in our two implementations: \textbf{(a)} the GPS version and \textbf{(b)} the triggered version.
    The essential difference is the method used for synchronising the verifiers' questions.
    In \textbf{(a)} the connection is wireless as it uses communication with satellites at the expense of a higher imprecision thus further verifier-prover pairs.
    In \textbf{(b)} the connection is physical and oriented from the first to the second verifier; the former sends a trigger through the fibre and delays their action by the time needed for this signal to reach the latter.
    With a better accuracy this second method allows for shorter distances between the verifier-prover pairs, here \SI{60}{\meter} but arguably improvable.
  }
  \label{fig:hardware}
\end{figure*}

\subsection{Hardware}
\label{app:hardware}

For the implementation, the verifiers consist of a standard computer (Intel core i3 processor with \SI{4}{\giga\byte} of RAM) and an FPGA development board (Xilinx SP605 evaluation board featuring Spartan-6 XC6SLX45T), the two being connected together through a PCI-Express link; the provers consist only of the same FPGA development board.
Within each verifier-prover pair, FPGA boards are communicating with each other through a \SI[per-mode=symbol]{2.5}{\giga\bit\per\second} small form-factor pluggable (SFP) optical link.
On the provers' side the main data (graph, colouring) is stored in memories available in the FPGA (Block RAM of about \SI{2}{\mega\bit}) and the random data on Flash memories available on the FPGA development board (\SI{32}{\mega\byte}), the latter being slower than the former.
This shared randomness was generated by means of the quantum random number generator (QRNG) Quantis by IDQuantique.

\subsubsection{GPS version}

A schematic view of the setup in this case is depicted on Fig.~\ref{fig:hardware}(a).
The verifiers' FPGAs are synchronised to the Coordinated Universal Time (UTC) by means of a Global Positioning System (GPS) clock, that is, a GPS receiver and an Oven-Controlled Quartz-Crystal Oscillator (OCXO) that creates a sinusoidal wave with a frequency of \SI{10}{\mega\hertz}.
This OCXO signal, locked to an electronic pulse per second (PPS) delivered by the GPS with a precision of \SI{150}{\nano\second}, is sent to the verifiers' FPGAs where its frequency is multiplied to a \SI{125}{\mega\hertz} signal through a phase-locked loop.
Eventually this \SI{125}{\mega\hertz} signal is used as a time reference for the computations performed on the FPGAs, which also receive the PPS signal to check the synchronisation with the GPS clock.
Specifically, we verified that there were $1.25\times10^6\pm1$ cycles between two successive PPS signal, fixing the cycle duration to \SI{8}{\nano\second}.
This shows that the inaccuracy added by the generation of the \SI{125}{\mega\hertz} clock would be below \SI{24}{\nano\second}.
Therefore, since the PPS signals are also labelled with a universal time stamp, the verifiers are able to synchronise their questions with an accuracy of $150+24=\,$\SI{174}{\nano\second}.

\subsubsection{Triggered version}

A schematic view of the setup in this case is depicted on Fig.~\ref{fig:hardware}(b).
The verifiers FPGA's are connected to one another with an SFP optical link, this link is used to synchronise the questions sent to the provers.
The FPGAs run at a base clock frequency of \SI{125}{\mega\hertz}.
The first verifier generates a stream of triggers at a rate of about \SI{3}{\mega\hertz}.
These impulsions are transferred to the second verifier through a fibre channel of \SI{62}{\meter} connecting both devices, in order to trigger the challenges sent to the prover.
On the first verifier this trigger is delayed by \SI{440}{\nano\second} to compensate for the delay in the fibre and the latency of the electronics.
With an oscilloscope we measured that the imprecision between the delayed trigger and the trigger sent through the optical fibre does not exceed three cycles, i.e., \SI{24}{\nano\second}.
Moreover the total time of the exchange in the verifiers' FPGA is inferior to 35 cycles but we determined that the verifiers internal latency, that is, the time when the data is still in the FPGA plus the time the answer is already back, accounts to at least 14 cycles, thus reducing this time to 21~cycles.
Note that this time arises from the conversion from electronic to optical signals.
When adding the imprecision of the trigger, we get that a round is achieved in a maximum of 24 cycles, that is, \SI{192}{\nano\second}, thus constraining the verifier to be at a minimal distance of \SI{57.6}{\meter}.

\subsection{Shared randomness}
\label{app:randomness}

In the protocol presented in Methods~\ref{app:protocol} the two provers need to have access to a common source of random information that they can use to share, in each round, their random colouring and randomisers.
For speed mostly (but also simplicity and elegance), it is preferable to have them \emph{store} this shared randomness.
Given the high number of rounds needed to reach a satisfying security and the relatively low memory of the FPGAs, a frugal approach is mandatory.
In this section we give the details of our implementation with this regard.

For the colouring, it is easy to see that there is a thrifty option: storing a fixed one and only drawing a random permutation of the colours.
The ``randomness cost'' of this part is therefore of one bit and one trit in each round.

For the randomisers, a naive approach would demand one random trit per vertex each time, thus requiring far too much memory given the high number of rounds.
Noting that only four of them are actually used in each round (two per prover), we apply a radically more affordable alternative: storing $|V|$ (the number of vertices) fixed trits and drawing $2m+1$ trits, where $m$ is the number of digits of $|V|$ in base three.
The idea is to expand randomness by assigning in advance a small ternary vector (of $2m+1$ trits) to each node; then each randomiser is simply chosen by computing the scalar product with a common random ternary vector (of $2m+1$ trits).
The subtle point to take care of is the independence of the resulting random variables, which amounts to the linear independence of the vectors assigned to the nodes.
As only four randomisers are consumed in each round, we want all sets of four such vectors to be free.
The literature luckily offers an elegant solution to this problem via ternary cyclic linear codes with minimum Hamming distance of five~\cite{LLH+14}.
The parity check matrix of a linear code with minimum Hamming distance $d$ is indeed such that all sets of $d-1$ of its columns are linearly independent; see, e.g., Ref.~\cite[Lemma 3.5]{TV09}.
Moreover, the cyclicity of the code used allows to store only one trit per node and to use it together with the ones of the next $2m$ nodes (in numerical order) to create its ternary vector.

\subsection{Quantum provers}
\label{app:quantum}

In this section we describe two adaptations of our protocol that are provably secure against quantum provers: one with a third verifier-prover pair~\cite{CL17} and one extending the size of the graph under study~\cite{Ji13}.
At the moment, these adaptations are not amenable to an experiment like ours; we discuss this point more precisely below.

Already in Ref.~\cite{CMS+19} the possibility of extending the protocol therein to three verifier-prover pairs following Ref.~\cite{CL17} is investigated.
Here we summarise the arguments, which are essentially similar for our simplified protocol.
The idea is to involve a third prover, also relativistically separated from the other two, and whose task will be to mimic one of them, randomly chosen by the verifiers.
It is easy to see that honest provers have no problem to pass this new version, as the added prover can simply share the same labellings and use them to answer exactly as before.
For dishonest provers trying to beat the protocol by means of quantum resources, the crucial point that prevents them from cheating rests on the monogamy of entanglement, namely, a fundamental trade-off among the amounts of entanglement a quantum system can have with others.

Unfortunately, the number of rounds for which security can \emph{currently} be proven for this protocol is about \red{$(11|E|)^4k$}, which is completely unpractical for graphs of reasonable size: \red{$2\times10^{18}$} rounds in our case, thus taking millennia!
Also we should mention that, with three provers, up to six vertices can be unveiled by the verifiers in every round, so that the storage of shared randomness would need a ternary code with a minimum Hamming distance of seven, for which the literature does not provide a solution as elegant as in Methods~\ref{app:randomness}.

Another alternative can be found in Ref.~\cite{Ji13}, where Ji suggests to inflate the graph into a bigger one for which security against quantum provers can be demonstrated.
The way to extend the graph is schematically the following: for all pairs of nonadjacent vertices, add four vertices following a certain pattern that Ji calls a \emph{commutative gadget}.
This little subgraph is indeed such that it enforces the strategy to involve commuting variables, and thus to be classical.

However, even though the increase is now only quadratic, classical and quantum security are not linked in Ref.~\cite{Ji13} so that the number of rounds required remains unknown.
Note also that memory may become an issue if the graph is too large, as this impacts not only the number of rounds, but also the amount of memory required in each round.

In summary, the adaptations of the protocol mentioned above are at the moment unable to provide a practical protocol sound against quantum provers.
Theoretical improvements following one of these directions, a combination of them, or a new one, would then be desirable in order to bring the number of rounds down to something practical.

\end{document}